\documentclass[article]{aa}
\usepackage{rotating}
\usepackage{longtable}
\usepackage{subfigure}
\usepackage{amssymb}
 \usepackage{graphicx}
 \usepackage{astro_bib_macro}
 \usepackage[authoryear]{natbib}
 \begin{document}
  \titlerunning{Mid-InfraRed imaging of Herbig Ae stars}
 \title{Mid-InfraRed imaging of the circumstellar dust 
around three Herbig Ae stars : HD~135344, CQ~Tau, HD~163296
\thanks{Based on observations obtained at the Canada France Hawaii Telescope 
(CFHT) which is operated by the national Research Council of Canada, 
the Institut National des Sciences de l'Univers of the Centre National 
de la Recherche Scientifique of France, and the University of Hawaii.}
} 
 \author{C. Doucet\inst{1} \and E. Pantin\inst{1} \and P.O. Lagage\inst{1} \and C. P. Dullemond\inst{2}}
 
 \institute{AIM, Unit\'e Mixte de Recherche CEA - CNRS - Universit\'e Paris VII - UMR 7158,
\\  DSM/DAPNIA/Service d'Astrophysique, CEA/Saclay, F-91191 Gif-sur-Yvette, France \and
 Max-Planck-Institut fur Astronomie Heidelberg, Konigstuhl 17, Heidelberg, Germany}
 \offprints{C. Doucet,\\
\email{doucetc@cea.fr}}

 \abstract
 {}
 {Planet formation has been known for many years to be tied
   to the spatial distribution of gas and dust in disks around young
   stars. To constrain planet formation models, imaging observations
   of protoplanetary disks are required. }
   { In this framework, we have
   undertaken a mid-infrared imaging survey of Herbig Ae
    stars, which are pre-main sequence stars of intermediate mass still surrounded 
 by a large amount of circumstellar material. The observations
   were made at a wavelength of 20.5 $\mu$m with the CAMIRAS camera
   mounted at the Cassegrain focus of the Canada France Hawaii
   Telescope.}
 { We report the observations of three stars, HD135344, CQTau and HD163296. The circumstellar material around the three objects is spatially resolved.  
 The extensions feature a disk like shape. The images provide direct information on two key parameters of the disk : 
 its inclination and its outer radius. The outer radius is found to be quite different from the one deduced from 
 disk models only constrained by fitting the Spectral Energy Distribution of the object. 
 Other parameters of the disk, such as flaring, dust mass have been deduced from fitting both the  
   observed extension and the spectral energy distribution with sophisticated disk models. }
   {Our results show how important imaging data
    are to tighten constraints on the disk model parameters.}
   


\maketitle 
 \section{Introduction}
     
The formation of circumstellar disks is a natural outcome of the star
formation process by which a molecular core collapses to form a star
\citep{SHU87}.  Circumstellar disks can outlive the
period during which stars form and still be present when the star is
in its Pre-Main-Sequence (PMS) phase. In these disks, composed of gas
and dust, various physical processes can lead to the growth of dust
grains and eventually to the formation of planets.  Understanding the
physical conditions that prevail in these objects is of crucial importance when
studying planet formation.\\ 
The study of circumstellar disks is a
field in fast development both from the observational and the modeling
point of view (e.g. \citet{NATTA04} and references
there-in). The Infrared Space Observatory (ISO) has given clues on the
dust composition of a sample of isolated HAeBe systems \citep{BOUW01, MEEUS01}.  While these spectra reveal the
composition of the dust, no direct information concerning the spatial
distribution of the different dust species can be inferred from the
ISO data. Most studies so far have used the Spectral Energy
Distributions (SEDs) to put constraints on the spatial distribution of
the circumstellar material.


Models of protoplanetary disks are increasingly successful at accounting for much of the
observed properties. For instance, they can justify that disks SEDs
are generally rather flat in $\nu F_{\nu}$, where $\nu$ is the
frequency and $F_{\nu}$ the flux \citep{KEN87}. Furthermore, models can explain that dust
features are almost all seen in emission (\citet{CAL91}, \citet{CG97}, hereafter CG97), the
presence of a near-infrared excess in the SEDs of Herbig Ae stars
(\citet{NATTA01}, \citet{DDN01}, hereafter DDN01) and interpret the
differences observed in the far-IR excesses \citep{DD02, DD04a}. Fitting the SED only allow to make conjectures
on how the disks' geometry looks like; spatially resolved imaging data of those disks are absolutely
necessary to verify theories and models' assumptions.
For instance, key parameters,
such as the disk surface density profile with radius, are still very poor
constrained when fitting SEDs.\\ 
Mid-infrared imaging observations from a large ground-based telescope are potentially well suited to
bring spatial information on disk around Herbig Ae (HAe)
stars. HAe stars represent the middle stage of PMS evolution of
intermediate-mass stars ($\sim$ 2-3 $M_{\sun}$); they are bright
enough to heat sub-micron dust grains at 100 AU to a temperature of
about 150 K.
Grains at such a temperature have their peak of thermal emission in
the mid-InfraRed (mid-IR). The diffraction limited angular resolution
achievable with a 3.6 meters class telescope in the mid-IR, 0.6/1.2
arcsec at 10/20 $\mu$m, corresponds to a distance of 60/120 AU for a
star located at a typical distance of 100 pc and thus allows a
 relatively good sampling of disks whose sizes range in the several
 hundreds of AU. The first attempts to resolve the spatial structure 
 of the circumstellar material around HAe stars were performed with multi-aperture observations with a single 
 bolometer; emission extending up to large distances from the star (more than 5 arcsec) were found 
 around 3 objects HD97048, HD97300 and HD176386 \citep{PRUSTI94}. 
 Such extended emissions were attributed to the emission from large 
 molecules and small grains transiently heated by star-light and distributed in a dust shell with a large inner radius, probably a remnant from the cloud from which the star was born. 
 It is only with the advent of mid-IR cameras that the full potential of mid-IR observations to 
 study disk structures was achieved. Single dish observations have revealed the disk structure in the 500 AU (Astronomical Units) range 
 around two HAe stars: AB Aur \citep{MAR95, PAN05} and HD100546 \citep{GRA01, LIU03}.
 Interferometric observations have allowed to probe the innermost regions of the disk (1-10 AU) and numerous objects have 
 been observed \citep{MILLAN01, TUTHILL02, Wilkin, LEINERT04, LIU05}. When mid-IR single dish observations have now clearly demonstrated on a few examples that they can provide unique information on 
 the disk structure, the constrains brought on the disk modeling by such spatial information has not yet been fully exploited.
 
 In this paper, we were interested in both increasing the number of HAe for which information on the mid-IR spatial 
 extension is available, and in using this information, in combination with the already existing SED 
 measurements between 1 and 100 $\mu$m, to constrain the parameters of the disk models 
 developed these last years. The paper is organized as follows :
 the observations and data reduction are described in Sect. 2. Sect. 3 deals with
 the results, in terms of spatial extensions. In Sect. 4, a first, simple approach is used to establish the disk inclination, and in
  Sect. 5, we describe the 2-D radiative transfer code used to reproduce the observations.
 A discussion of the results follows in Sect. 6.
 Conclusions and perspective are drawn in Sect. 7.
\section{Observations and data reduction}
We have observed a sample of three Herbig Ae stars: HD~135344, CQ~Tau and HD~163296. 
Table~\ref{objets} presents the main stellar parameters of the sample. The objects were
selected from the catalogue of \citet{THE94} and \citet{MALF98} according to the following criteria: the objects are bright in the mid-IR, relatively close and isolated, 
i.e not associated to extended diffuse emission due to the parental cloud. 
 
CQ~Tau is located at a distance of 100$_{-17}^{+25}$ pc and has an age of 10 Myr 
\citep{NATTA01, VDA98}. HD~163296 is at a distance 
of 122$_{-16}^{+13}$ pc and has an age of 7 $\pm$ 5 Myr \citep{VB05}.
The distance and age of HD~135344 are more controversial. Until 2001, a distance of 84 pc 
\citep{MEEUS01} with an age of 17 $\pm$ 3 Myr \citep{THI01} were used 
for this object. But, in a recent paper \citep{VB05}, the distance 
was re-evaluated to 140$\pm$42 pc and the age to 
8 $\pm$ 4 Myr; these latter values will be used in the following.
  
The observations were performed with the mid-IR camera CAMIRAS
\citep{LAG92} installed as a visiting instrument at the
Cassegrain focus of the Canada France Hawaii Telescope (CFHT). The
camera is equipped with a Boeing 128x128 pixels Blocked Impurity Band
(BIB) detector sensitive up to a wavelength, $\lambda$, of $\sim$ 27
$\mu$m. A
filter centered at 20.5 $\mu$m with a Full Width Half Maximum (FWHM) bandpass,
$\Delta \lambda$, of  1.11 $\mu$m was used. The Pixel Field of View (PFoV) on the sky was 
0.29 arcsec ; such a PFoV provides a good sampling 
of the diffraction pattern which is of 1.5 arcsec FWHM.\\ 
The objects were observed between 2000 March, 18$^{th}$ and 2000 March, 24$^{th}$.  
During the run, seeing and weather conditions were extremely
favorable and stable in time. HD~163296 and CQ~Tau were observed at
a median airmass of 1.4, and HD~135344 at an airmass around 1.9, which is the best 
achievable when observing from CFHT. 
Standard chopping and
nodding techniques were applied to suppress atmosphere and telescope background emissions; 
the chopping throw was 16 arcsec to the North and the frequency used was 3.33 Hz; 
the nodding amplitude was 20 arcsec to the West. 
The nodding direction was perpendicular to
the chopping direction, in order to get the best spatial
resolution; given the low chopping and nodding throw and the field 
of view of the camera, 
the source always remained within the detector field of view, the obtained images contain thus 4 beams (2 positive, 2 negative). Given the huge photon background in the mid-IR, the elementary integration time was set 
to 15 ms, and the images were co-added in real time
 in order to store only two co-added images (one for each chopping position) every second.\\
\begin{table*}[ht]
\caption{Properties of the three Herbig Ae stars observed with the CAMIRAS mid-InfraRed camera; 
first two columns: coordinates of the objects (right ascension and declination); third column: 
their spectral type; fourth column : their distance, as deduced from Hipparcos data; fifth column: 
flux in the IRAS 25 $\mu$m band; sixth column : flux at 20 $\mu$m obtained 
in this paper; seventh column : age of the objects in Myrs.
{\it References: (1) \citet{COUL98}, (2) \citet{DD03}, (3) \citet{JAYA01}, (4) \citet{MEEUS01}, (5) \citet{MANSARG97}, (6) \citet{MANSARG00}, (7) \citet{NATTA01}, (8) \citet{SYL96}, (9) \citet{TESTI01}, (10) \citet{THI01}, (11) \citet{VB05}, (12) \citet{VDA98}.}}\label{objets}

\begin{tabular}{lllcccccl}
 \hline\hline
 Object & RA (2000) & Dec (2000) & Spectral type &Distance (pc) & F$_{25}$ (Jy) & F$_{20}$ (Jy)& age (Myrs)& References\\
 \hline
HD 135344 & 15 15 48.4 & - 37 09 16 & F4V& 140$_{-42}^{+42}$& 6.7 & 5 $\pm$ 1   &8$\pm$4 &\tiny{1,2,4,8,10,11}\\
HD 163296 & 17 56 21.4 & - 21 57 20 & A3Ve & 122$_{-16}^{+13}$ & 21 &18 $\pm$ 4   & 7$\pm$5 & \tiny{3,4,5,11,12}\\
CQ~Tau & 05 35 58.4 & +24 44 54 &   A1-F5IVe &   100$_{-17}^{+25}$& 20.6 & 23 $\pm$ 3 & 10& \tiny{6,7,9,12}\\
\hline
\end{tabular}
\end{table*}

\begin{figure*}
\vbox{
\centerline{
\hbox{
\resizebox{8.7cm}{!}{\includegraphics[angle=90]{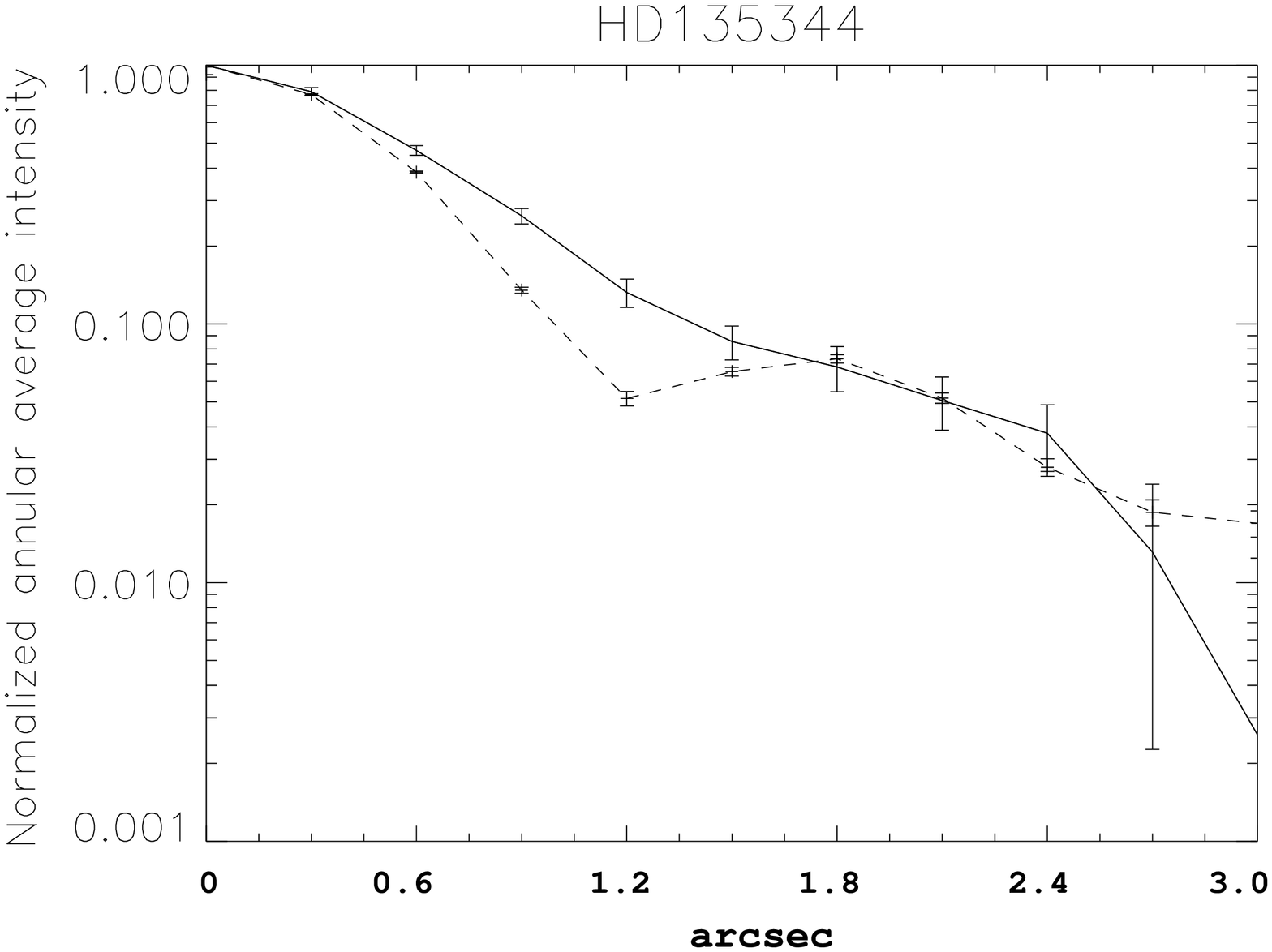}}
\resizebox{8.7cm}{!}{\includegraphics[angle=90]{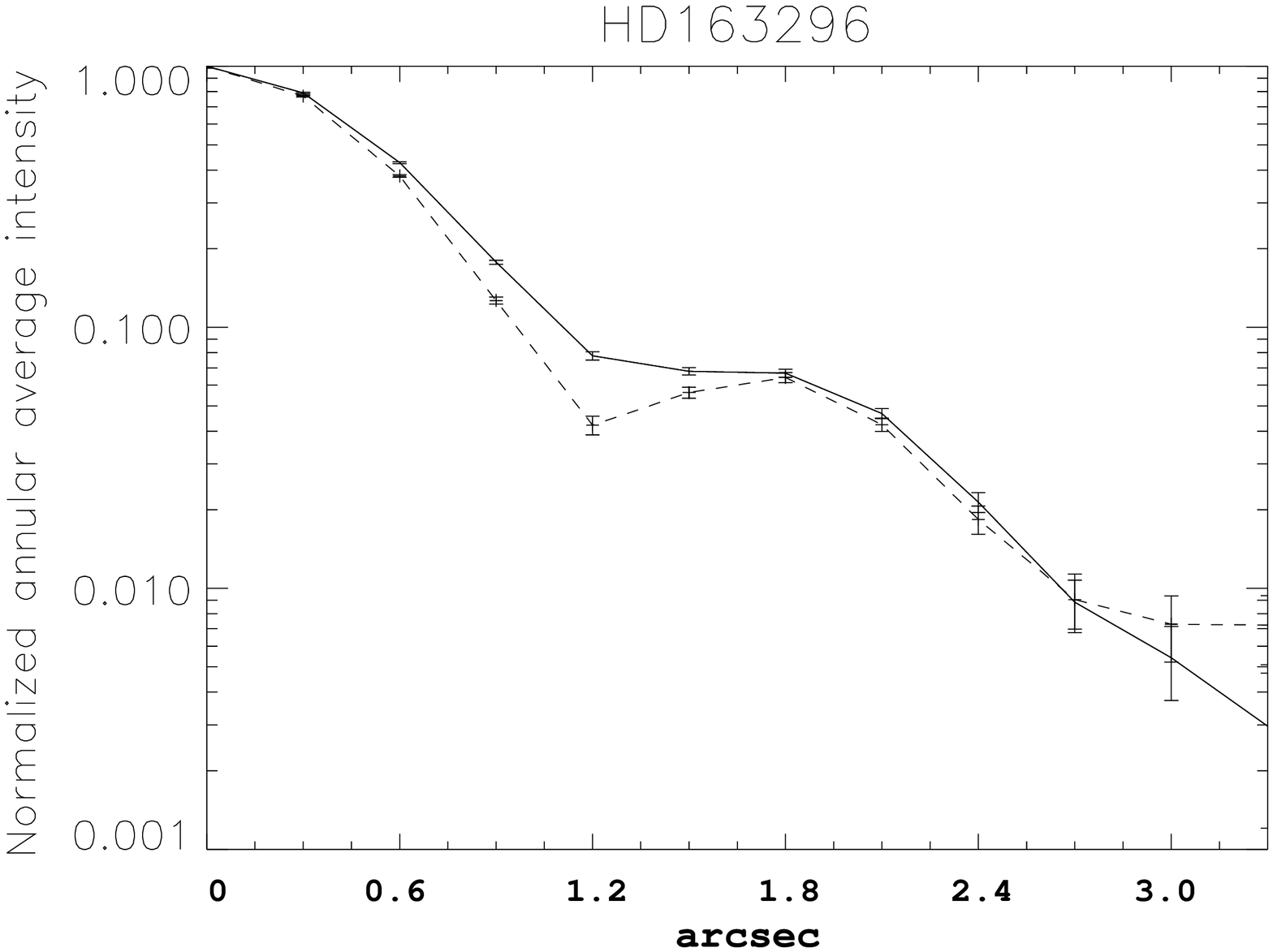}}
}}
\centerline{
\hbox{
\resizebox{8.7cm}{!}{\includegraphics[angle=90]{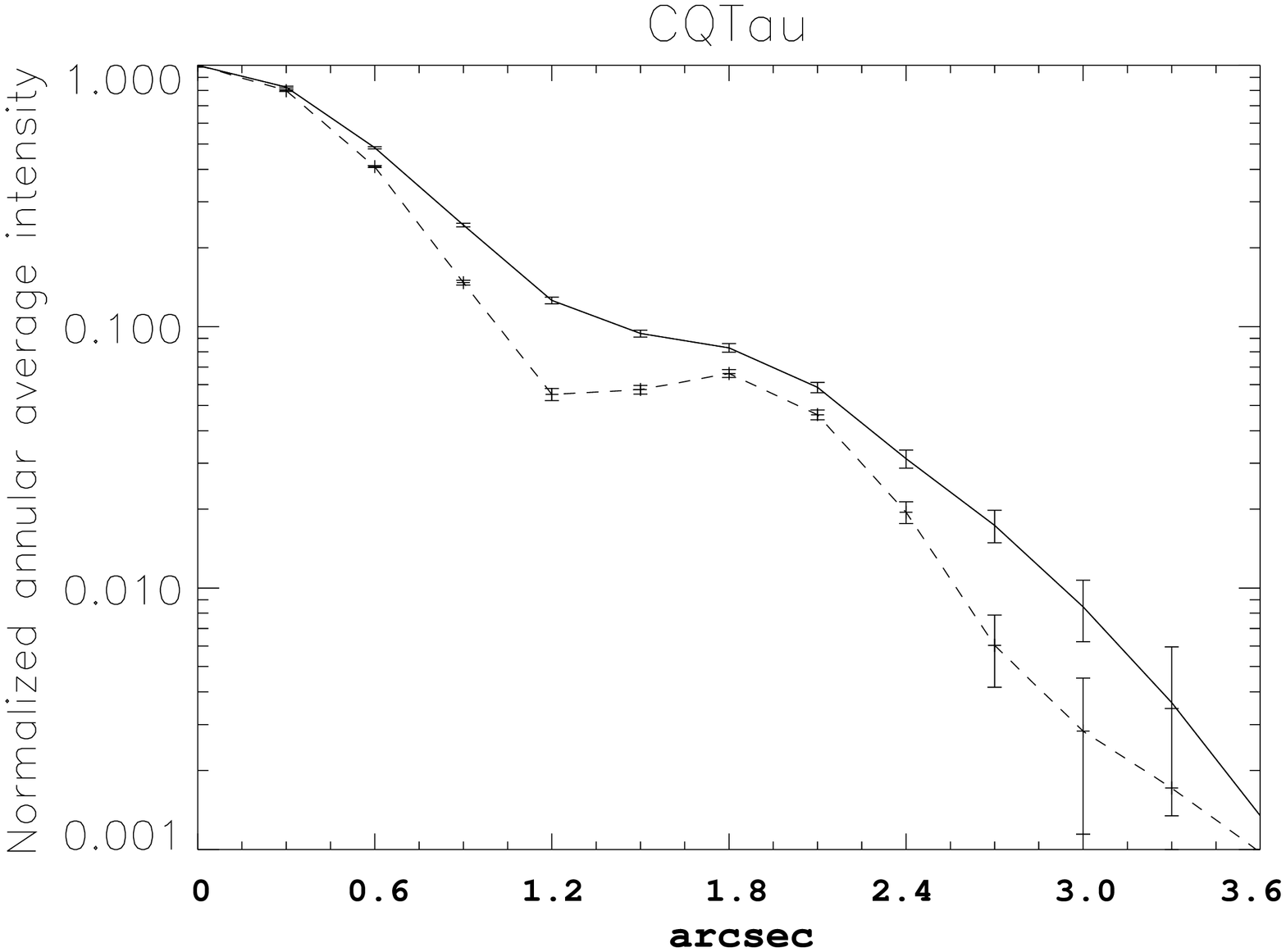}}
\resizebox{8.7cm}{!}{\includegraphics[angle=90]{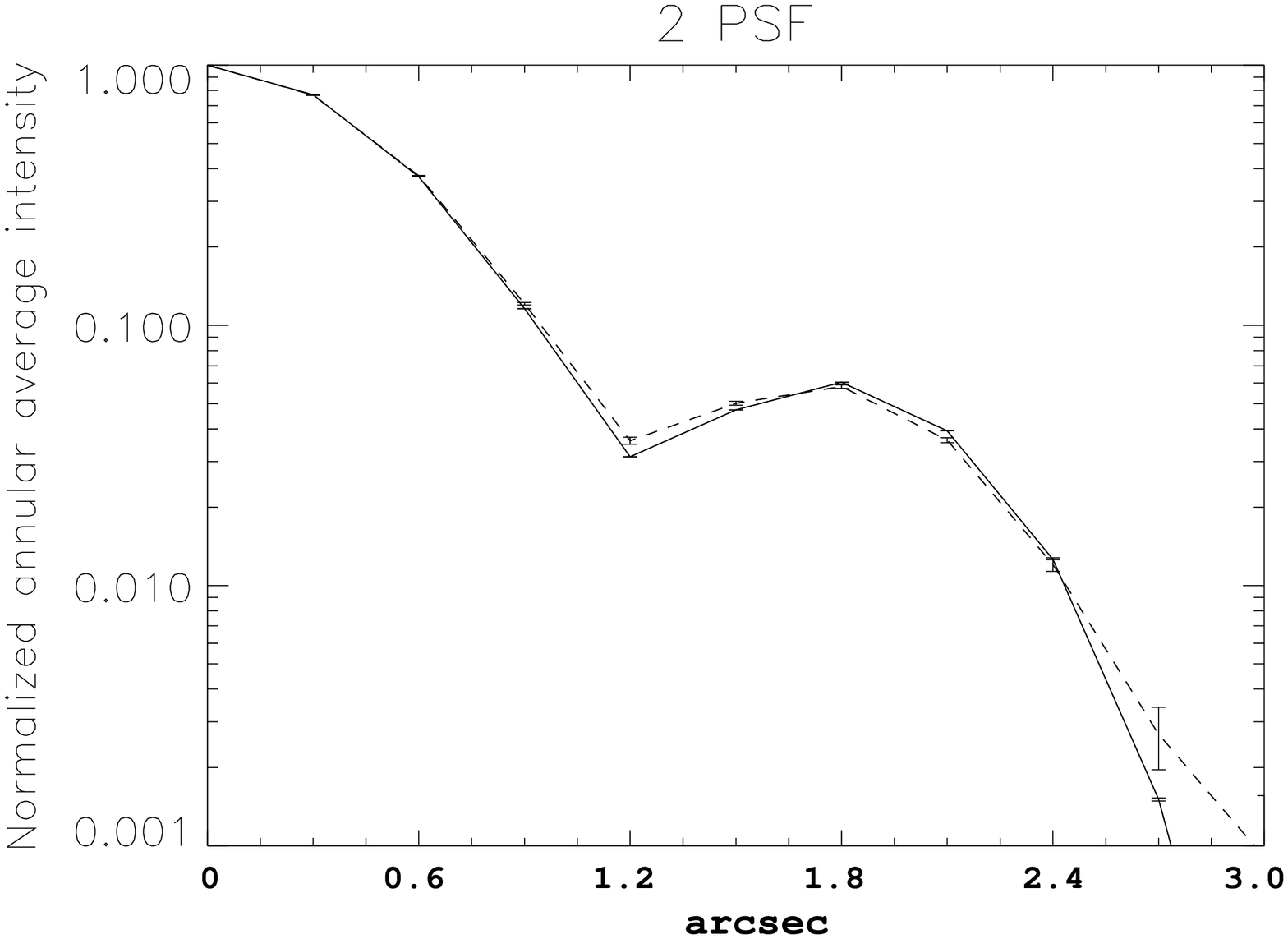}}
}}
}
\caption{Annular averages intensity aperture for the source (solid
  line) and PSF (dashed line) normalized to the peak value, as
  function of radius. The error drawn are 1 $\sigma$-errors RMS. The reference star for
  HD~135344 is $\alpha$ Boo, for HD~163296 $\gamma$ Dra and for CQ~Tau,
  $\beta$ Gem. The plot labeled '2 PSF' shows two reference stars
  $\alpha$ Boo and $\gamma$ Dra as observed on the 24$^{th}$ of March
  2000. Although the two reference stars have very different fluxes
  ($\gamma$Dra with 43 Jy and $\alpha$ Boo with 170 Jy, \citet{VANMALDEREN}), the two
  profiles are similar.}

\label{oplot_cuts}

\end{figure*}
 
  The basic data reduction is standard. The data cubes of one observation 
are carefully stacked with rejection of corrupted planes. A shift-and-add procedure is applied to each cube of
 images using a correlation based method with a re-sampling factor of 8:1. 
The four beams are then combined
 in one image by a source extraction algorithm followed by a shift-and-add procedure.
 Finally, flux calibration is achieved via aperture photometry of a set of 
photometric standard stars such as $\alpha$Tau, $\alpha$Boo, $\beta$Gem or $\gamma$Dra 
\citep{COHEN99}. The photometry gives a total flux of 6.3$\pm$0.6 Jy for HD~135344, 
23$\pm$3 Jy for CQ~Tau and 18$\pm$4 Jy for HD~163296, in good agreement with the IRAS values 
(Tab.~\ref{objets}).

 \section{Extended emission}
  
The three objects of the sample are spatially resolved. This can be seen on 
Fig.~\ref{oplot_cuts} where we have compared the average annular profile of the object with those of the Point Spread Function (PSF), obtained from the observation of a reference point-like star. 
Extended emission is detected up to 100-300 AU. 

We carefully checked that the observed extensions are not artifacts,
but the result of true extended emission from the objects.  Several
arguments lead us to reject explanations of the extensions in terms of
temporal variations of the PSF between the observation of the object
and the observation of the reference star. One possible cause of
such temporal PSF variations could be variations of the seeing. This
hypothesis is however rejected for the following reasons.  First,
concerning the limitations of the spatial resolution, the seeing
contribution at 20.5 $\mu$m is negligible with respect to the
diffraction : for a typical seeing value of 0.8" FWHM in the visible
range, one can estimate a seeing contribution at 20.5 $\mu$m around
0.4", when using the $\lambda^{(-1/5)}$ scaling law; thus, seeing
induces PSF changes of the order of 1 pixel FWHM, which is much
smaller than the widths of observed extensions.  Secondly, the seeing
was quite stable during the observations; thus we estimated seeing
variation effects to be much lower than one pixel.  Note also that two
of the objects were observed during different nights.  HD~135344 was
observed for four different nights: on the March, 18$^{th}$, 2000 (exposure time 6
mn), on the March, 19$^{th}$ (exposure time: 3 mn), on the March, 21$^{st}$
(exposure time: 6 mn) and on the March, 24$^{th}$ (exposure time: 2
mn 30). Its extension is confirmed over the 4 nights. CQ~Tau was
observed during 2 different nights on the 20$^{th}$ of March and on the 21$^{st}$ of
March (respectively with 3 and 6 mn of integration time) and is
spatially extended in both datasets.  HD~163296 was observed only once
(exposure time: 3 mn) on the 24$^{th}$ of March. 

Another possible source of fake extensions could be chromatic
effects. Indeed, HAe stars have large infrared excesses, and thus
have SEDs quite different from that of the PSF reference stars. Any
filter leak, either on the blue or on the red side of the nominal
filter bandpass, would then potentially lead to PSF variations between
point-like HAe objects and PSF reference stars. We double checked the
filter transmission at the operating temperature of the filter (10K)
using a Fourier Transform Spectrometer with a spectral resolution of 4
cm$^{-1}$.  The rejection rate outside the filter bandpass is
typically better than 10$^{-3}$.  We simulated the PSF variations due
to such a filter using the mid-IR spectra of the objects
obtained with ISO \citep{MEEUS01} for two of the three extended
objects; (CQ~Tau was not observed by ISO SWS).  The simulated PSF
variations lead to some extended emission, but much fainter than
observed, both in intensity and in spatial extension. Thus,
explanation of the observed extensions in terms of chromatic
variations of the PSF can be discarded.  \\ 

\begin{figure}[!h]
\resizebox{8cm}{!}{\includegraphics[scale=0.12,angle=90]{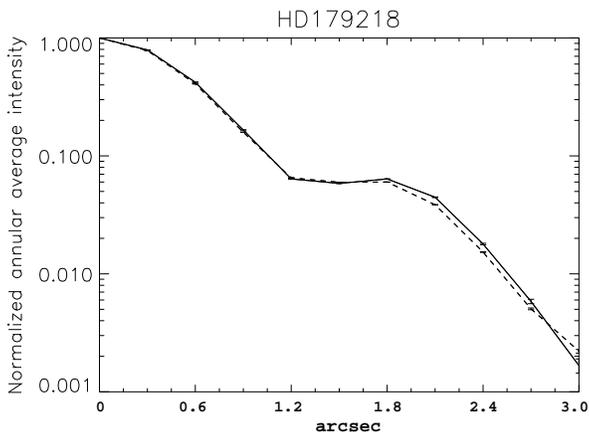}}
\caption{Same as Fig.~\ref{oplot_cuts}, to make a comparison with HD179218 for which there is no extended emission detected.}\label{HD179218}
 \end{figure}

Note also that not all the observed HAe stars observed are extended. In order to make comparison, a
far away HAe star, with a similar (even more) IR excess, HD179218 located at 240 pc \citep{MEEUS01}, was observed.  This object does not show extended emission at 20.5
$\mu$m (Fig.~\ref{HD179218}). This is an additional argument to
conclude that the extension observed in HD~135344, CQ~Tau, HD~163296 is
really due to an extended emission from the objects.



\section{Inclination of the disk}

Extensions around two out of the three objects, namely CQ~Tau and HD~163296, were
already observed at other wavelengths. \citet{TESTI01} have
resolved the emission around CQ~Tau at 7 mm and concluded that it was
compatible with a disk-like geometry.  \citet{GRA01} have
obtained coronographic images of HD~163296 with the Space Telescope
Imaging Spectrograph on board the \emph{Hubble Space Telescope} which
revealed a circumstellar disk with a radius of 450 AU. Therefore, our
modeling of the extensions seen in the mid-IR range assumes {\em
 a priori} a disk-like geometry.\\

\begin{figure}[!t]
 \resizebox{\hsize}{!}{\includegraphics[angle=0,scale=0.48]{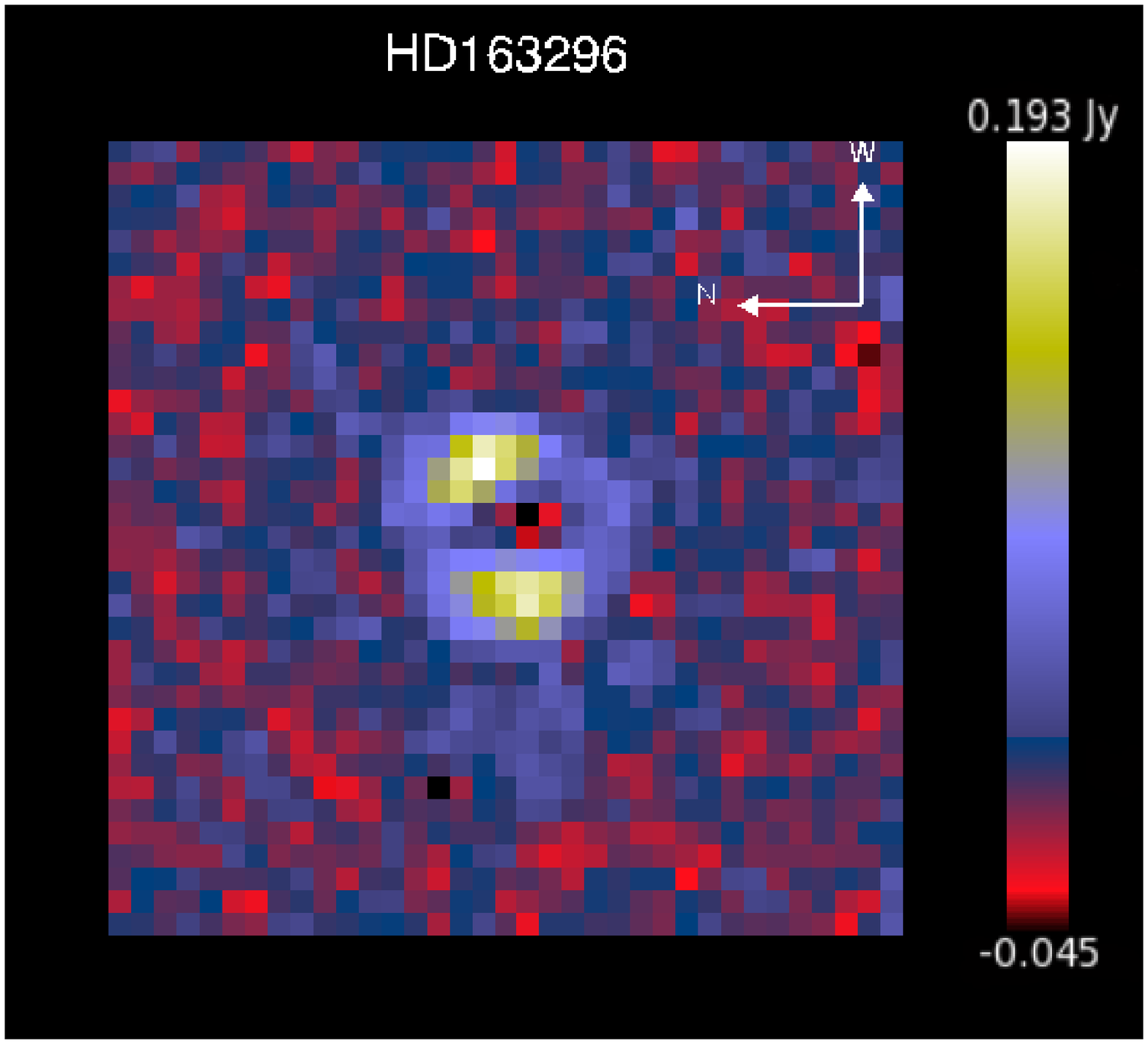}}
\caption{HD~163296 extended emission (central point source removed). West is up, North on the left. The pixel size is 0.29 arcsec. The disk has a surface brightness of 0.59 Jy/$^{"2}$ (S/N=27 for $\sigma \sim$ 0.002 Jy) for the brightness part in the direction east/west and 0.23 Jy/$^{"2}$ (S/N=11) for the less bright part in the direction north/south. The noise is calculated using a sigma-clipping technique. 
}\label{hd163296}
\end{figure}

\begin{figure}[!h]
 \resizebox{\hsize}{!}{\includegraphics[scale=0.1]{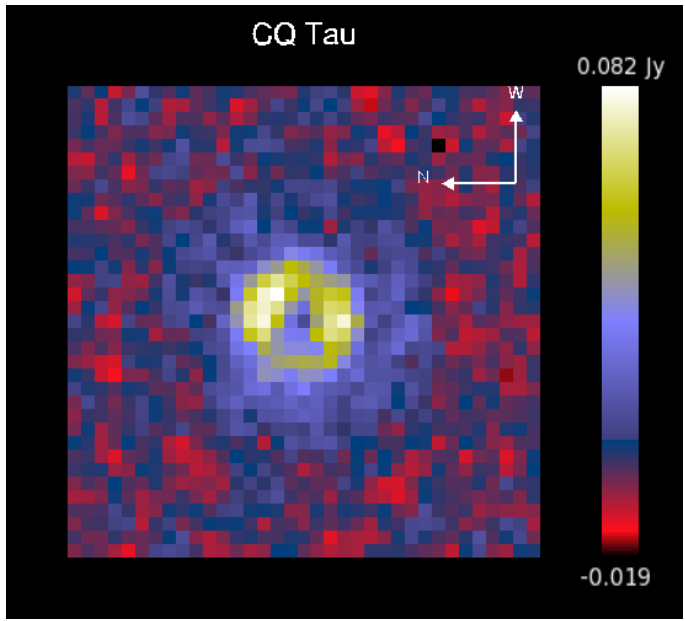}}
\caption{CQ~Tau extended emission (central point source removed). The same orientation as Fig~\ref{hd163296}. The pixel size is 0.29 arcsec. At 0.9" from the center, the surface brightness is 0.94 Jy/$^{"2}$ (S/N=15 for $\sigma \sim$ 0.004 Jy). At 2" from the center, the surface brightness is 0.23 Jy/$^{"2}$ (S/N=5). 
}\label{CQtau}
\end{figure}

\begin{figure}[!h]
 \resizebox{\hsize}{!}{\includegraphics[scale=0.5,angle=0]{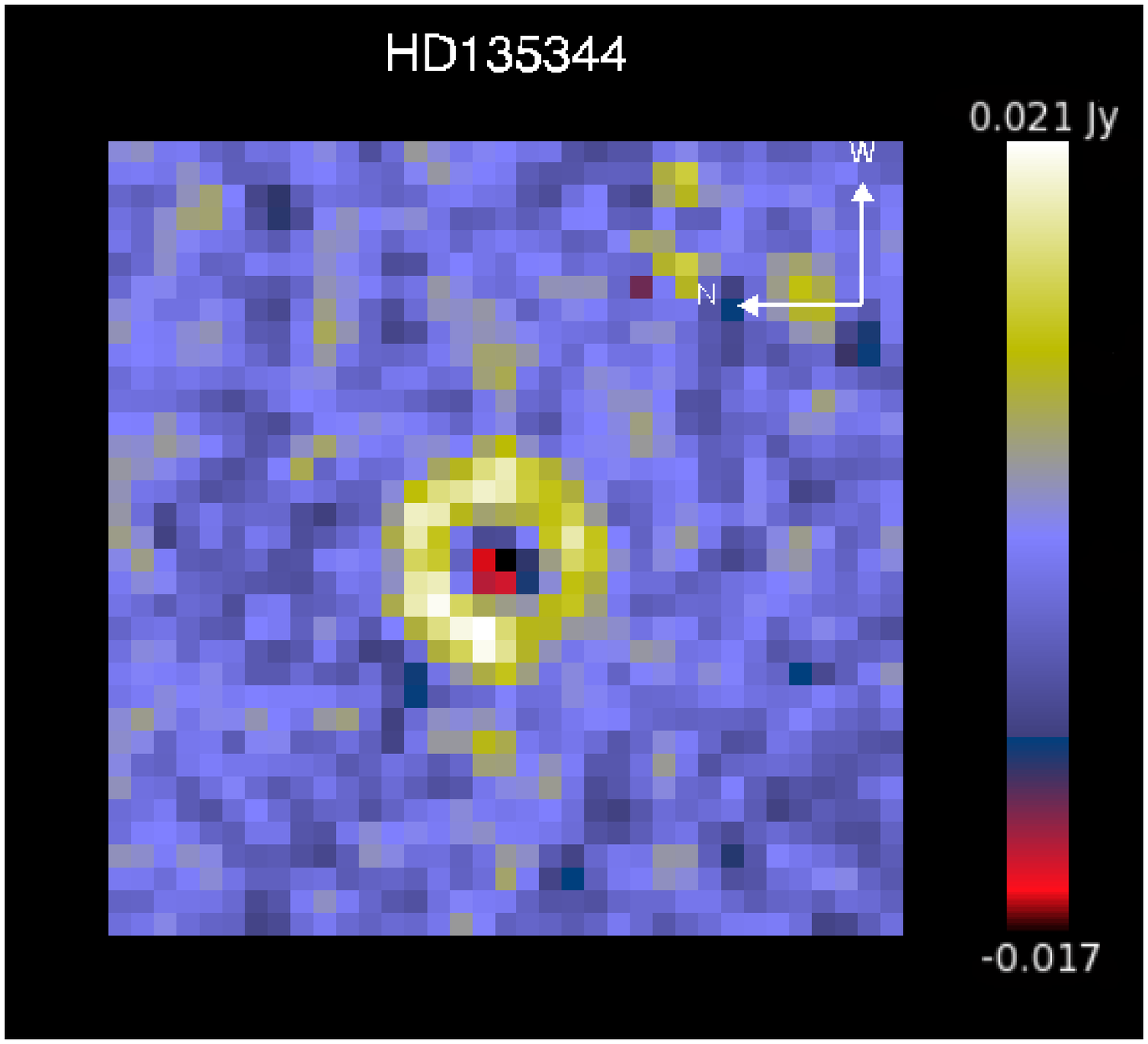}}
 \caption{HD~135344 extended emission (central point source removed).The same orientation as Fig~\ref{hd163296}. The pixel size is 0.29 arcsec. At 0.9" from the center, the surface brightness is 0.25 Jy/$^{"2}$ (S/N=15 for $\sigma \sim$ 0.0015 Jy). 
 }\label{hd135344}
 \end{figure}

Disks inclinations can be relatively easily determined if their
emission are spatially resolved. 
However, HAe disks are generally dominated in the mid-IR by the innermost regions (1-30 AU). Our goal was to detect the emission
from the \emph{intermediate} regions of the disks (30-200 AU). The disk emission can be decomposed, in our image, in a central 
\emph{unresolved} component plus an extended one (which geometry should reflect the true disk geometry at the distance scales achievable for our data). 
We first removed the "point-like"
central emission component by subtracting a scaled PSF to the image of
the object. The parameters of the point-like component (intensity,
position) were computed automatically using a penalty functional (in
order to avoid any visual bias) and then cross-checked visually using
a dedicated graphical interface built in IDL.  The resulting image,
called {\em residuals}, is free from the central emission, so 
that the extended emission is enhanced and it is easier to determine the disk geometry. 
For each target, this processing was done
using all available PSF measurements. The errors on the putative
extensions were assessed when applying the same procedure to two PSF
reference stars and, when possible, when comparing the extensions
obtained from different nights. The results are shown in
Fig.~\ref{hd163296},~\ref{CQtau} and~\ref{hd135344}.  \\

The first result is the elliptical shape of the extensions, which is
characteristic of the emission of disks inclined with respect to the line of sight.
 An ellipse fit of the residuals gives an estimate of the disks inclinations and
position angles.  
The results in terms of inclination and position angles are shown in Tab.~\ref{inclination}.
 
 The inclination of 33$^{\circ} \pm$ 5 found here for CQTau is in the middle range of values found in the literature, which range from $63^{\circ}$ \citep{TESTI01} to 14$^{\circ}$ \citep{DENT05}; in between we can find $48^{\circ}$ $^{+4^{\circ}}_{-3^{\circ}}$ ($PA = 105^{\circ} \pm 5$) \citep{EISN04}.
 
 


For HD~163296, the disk inclination of 60$^{\circ} \pm$ 5$^{\circ}$ is in good agreement 
with that found by Mannings \& Sargent (1997) of 58$^{\circ}$. 
Concerning the position angle of the disk, \citet{MANSARG97} 
found 126$^{\circ} \pm$ 3 with CO observations, \citet{GRA00} found 140$^{\circ} \pm$ 5 
thanks to optical coronographic images. We found a value of 105$^{\circ} \pm$ 10 when
fitting our mid-IR data. 
This difference in the position angle could be related to the fact that our data are only 
sensitive to warm dust of which geometry could slightly differ from that seen at shorter or longer
wavelengths, if the disk contains for instance non axisymmetric structures.

\begin{table}[!t]
\caption{Parameters of the elliptical fit of the image. In the first column, the inclination of the disk (0 degree means a faced-on disk). The position angle (PA) of the disk
on the sky is measured counter-clockwise from the North. The distance corresponds to the maximum distance at which the disk is detected in the 20 $\mu$m observations. The brightness level in the last column corresponds to the brightness of the maximum distance at which the disk is detected here.}\label{inclination}
\begin{minipage}{\hsize}
\setlength{\tabcolsep}{1.5mm}
\renewcommand{\arraystretch}{1.15}
\renewcommand{\footnoterule}{}
\begin{tabular}{lcccc}
 \hline\hline
          & i  & PA  & distance  & Brightness level\\  
& (deg)   & (deg) & (AU) & (mJy/"2)\\
\hline
HD 135344  &  46 $\pm$ 5 & 100 $\pm$ 10 & 210 &170\\
HD 163296  &  60 $\pm$ 5 & 105 $\pm$ 10 & 215 & 135 \\
CQ~Tau  & 33 $\pm$ 5 & 120 $\pm$ 10 & 290 & 129\\
\end{tabular}
\end{minipage}
\end{table}


\section{Modeling}

We have used a relatively simple parameterized model to investigate the dependence of
the emission in the mid-IR on each parameter. We consider disks heated by
irradiation from the central star. The
density profile of the gas is parameterized as a function of r (radius) and z (vertical height above the disk mid-plane):
\begin{equation}
\rho(r,z) = \frac{\Sigma(r)}{\sqrt{2\pi}H_p(r)}
\exp\left(-\frac{z^2}{2H_p(r)^2}\right)
\end{equation}
and it is assumed that the dust is well mixed with the gas. The surface
density is assumed to follow a power-law in radius: $\Sigma(r)=\Sigma_{0}
(r/r_{0})^{-p}$, with $r_{0}$ a fiducial radius. The scale height of the
disk $H_p(r)$ is also assumed to be a power-law: $H_p(r)=H_0(r/r_{0})^{q}$.
The inner radius of the disk (r=$R_{in}$) is located at the dust evaporation
radius (1400 - 1500K for silicate dust). 
The inner boundary (rim) is directly exposed to the stellar flux and is puffed up since it is hotter than the rest of the disk. 
Here, we mimic the puffing-up of the rim predicted by DDN01 by a specified value of $H_{p,\mathrm{in}}$ at $r=R_{in}$, which is a parameter of the model.
It should be noted that whether such an inner rim
is indeed puffed-up is still a matter of debate. Moreover, \citet{ISEL05} have shown that the rim is probably rounded-off due to the
density-dependence of the dust sublimation temperature. This effect is
not included here.


For the dust opacities, we use those of \citet{DLI01}. We
use a MRN \citep{MRN} distribution of grains
($n(a) \propto a^{-3.5}$) with a size between 0.01 and 0.3 $\mu$m. It is the disk surface layer which dominates the SED in the mid-IR range. The emission of the surface is made by small grains which trace the disk geometry. 
We will focus in this paper on this component.    
Since the objects have no PAH (Polycyclic Aromatic
Hydrocarbon) emission (or weak concerning HD~135344),
we do not take into account this population of grains.


Once the density profile is set, the dust opacities and stellar parameters
given, the code {\tt RADMC} \citep{DD04a, PON05} solves the temperature structure of the disk in a
Monte-Carlo way using a variant of the algorithm of Bjorkman \& Wood (1997).
This Monte-Carlo code also produces the source terms for scattering, in the
isotropic-scattering approximation.  With a ray-tracing tool (which is part
of the code {\tt RADICAL}, see \citet{DUTULL00} for a detailed description) the SED and
images can then be produced and compared to the observations. 
Comparative images are obtained by first resampling the maps to
CAMIRAS sampling and then by convolving them with the PSF.


\section{Comparison model versus observation}
\begin{figure}[ht]
 \begin{center}
 \resizebox{\hsize}{!}{\includegraphics[scale=0.1]{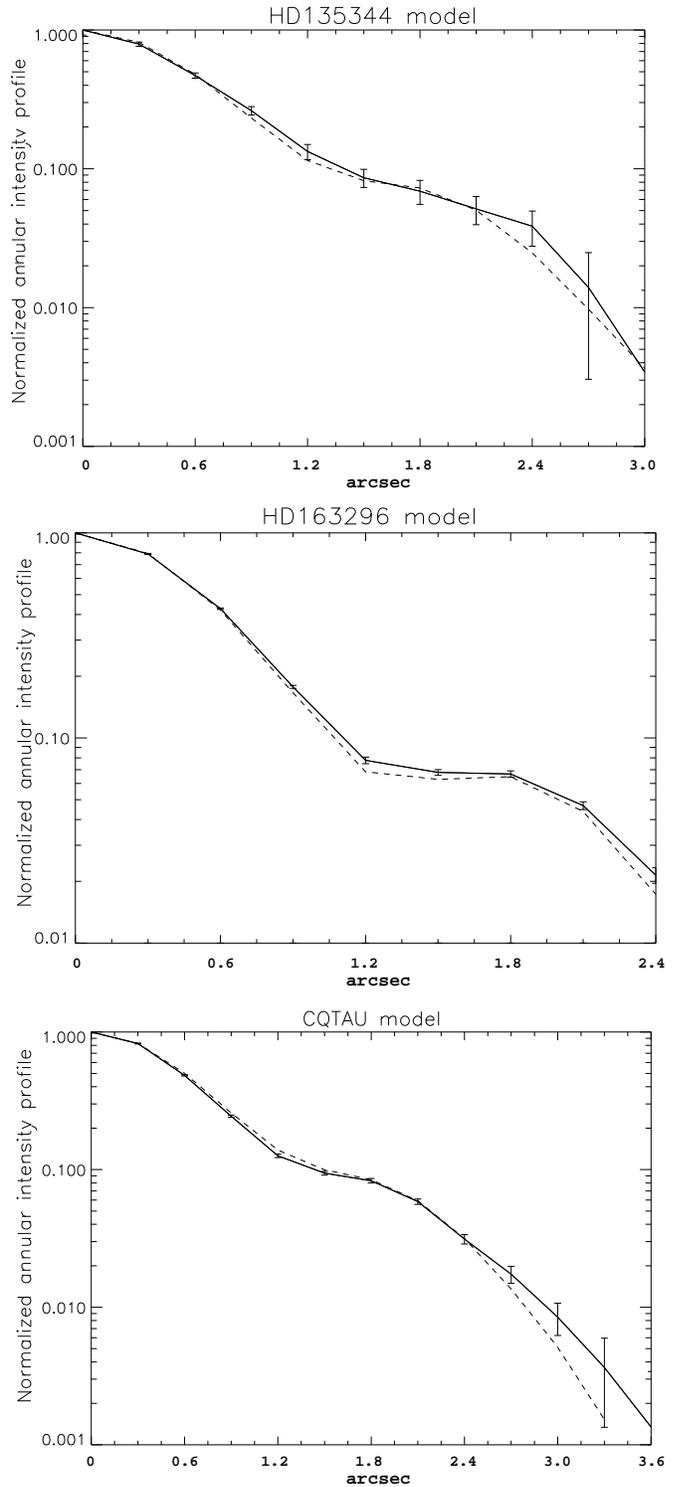}}
\caption{Annular average intensity aperture normalized to the peak for the observations (\emph{full line}), model (\emph{dashed line}) and the PSF associated (\emph{dot-dashed line}). We have constructed an image with the DD04 model that we have convolved with the PSF of the night associated. The error in the profile comes from the noise RMS in the image.}\label{model_profile}
\end{center}
\end{figure}

\begin{figure}[ht]
 \resizebox{\hsize}{!}{\includegraphics[scale=0.1]{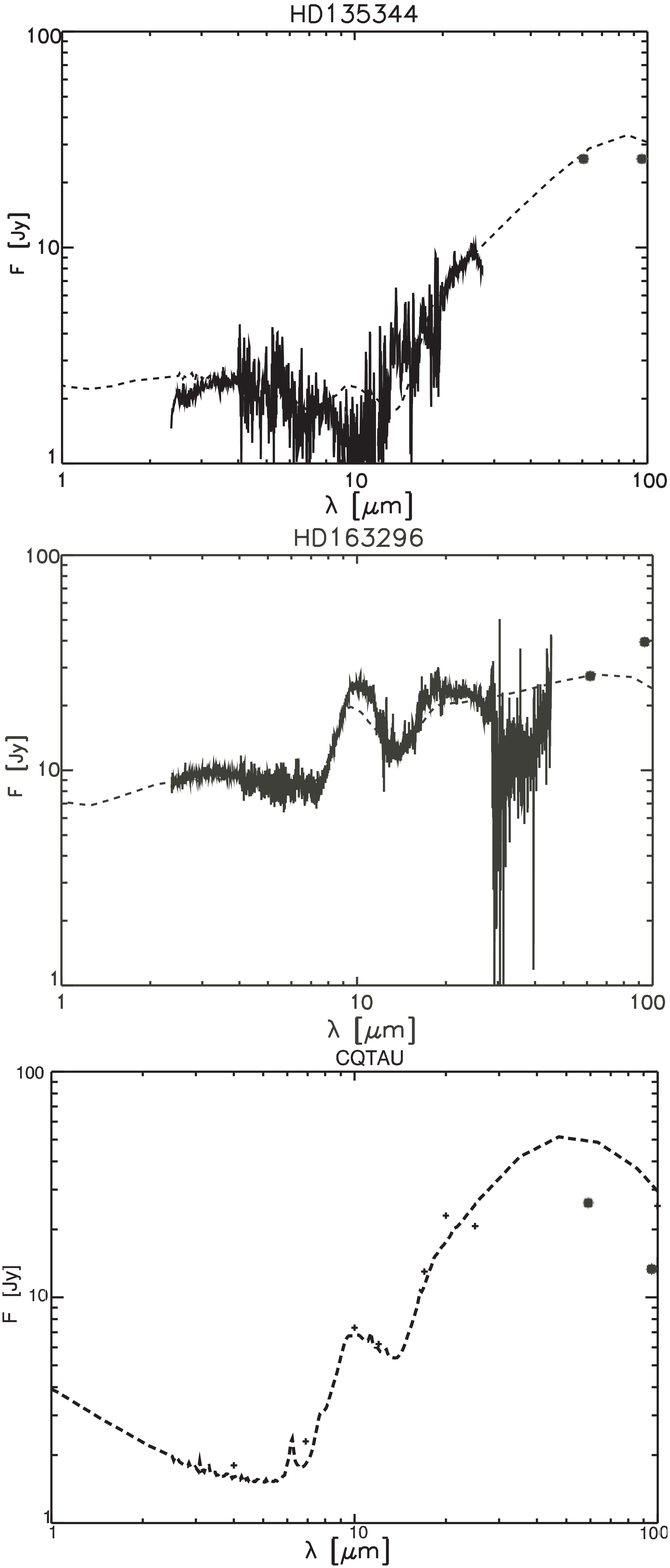}}
\caption{Modeled spectrum (dashed line) between 1 and 100 $\mu$m. For HD~135344 and HD~163296, the ISO spectrum is overploted in bold line. For CQ~Tau, there are some photometric points from\citet{GRA05}; we have added the one from this study at 20.5 $\mu$m and IRAS points at 60 and 100 $\mu$m (full circle).}\label{spectrum}
\end{figure}

\begin{figure}[ht]
 \begin{center}
 \resizebox{\hsize}{!}{\includegraphics[scale=0.1]{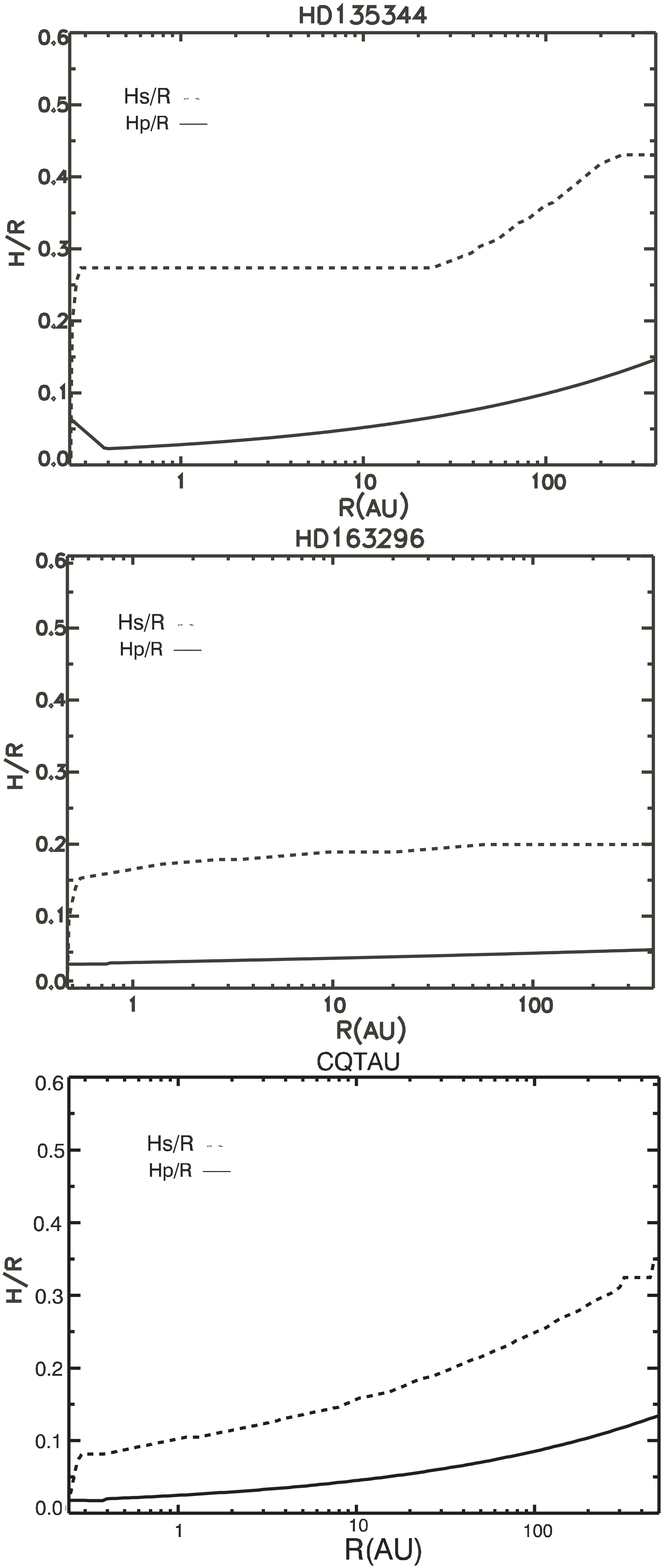}}
\caption{Modelized vertical structure versus the radius. The pressure height scale ($H_{p}/R$) is in full line and the surface height scale (i.e. the height of the disk photosphere above the mid-plane $H_{s}/R$) is the dashed line.}\label{structure}
\end{center}
\end{figure}

For each object, the best model shall fit simultaneously the SED (Fig.~\ref{spectrum}) and the extension found at 20.5 $\mu$m with our
observations (Fig.~\ref{model_profile}). Concerning HD~135344 and HD~163296, we used mainly the ISO spectrum to constrain the SED; as far as CQ~Tau is concerned,
IRAS photometry and BASS points obtained by \citet{GRA05} are used. Figure~\ref{structure} shows the structure of the disk in terms of pressure and surface scale height.
 
 Multiple runs of the model are performed until a satisfactory fit to the
observed spectrum and the extension found at 20.5 $\mu$m is obtained. In the fitting
procedure, the stellar parameters ($T_{eff}$, $M_{\star}$, $R_{\star}$,
see Tab.~\ref{resultat_model}), the dust evaporation temperature
(i.e. the position of the inner rim) at 1400 K, the outer radius and
the inclination of the disk, the dust composition and size distribution are fixed. 
Other parameters, such as the pressure scale height for the inner rim (eventually puffed-up) and the outer pressure scale height ($H_{0}$),
the power-law index of the pressure scale height (i.e. $q$, which has been fixed in the case of flared disk to 9/7, a value determinated by hydrostatic equilibrium \citet{CG97}), the mass of the
disk and the power-law index of the surface density (i.e. $p$) are estimated. Fitting the SED gives one solution among several degenerate combinations. 
A minimum value of the outer radius is derived from 20 $\mu$m observations, which is a very strong constrain on the true disk size, thus removing largely the degeneracy on the set of model parameters.

As first guesses, we used those parameters
found when fitting the SED by \citet{DD03} for HD~135344 and HD~163296 and by \citet{CHIANG01} for CQ~Tau.  
Trying to fit the SED, we focused
mainly on the Near-IR and mid-IR regions. Indeed, those regions are the regions
where most of the reprocessed stellar energy (by the disk's surface layer) emerged, and therefore the
most strongly affected by the model and the geometry of the disk.

 

Not all the parameters are sensitive to the spatial distribution at 20.5 $\mu$m (Tab.~\ref{influence_para}). The strong constrain that put the CAMIRAS images is the minimum outer radius, and we have to find a solution with all the free parameters to reproduce the shape of the SED, the extension and the total flux at 20.5 $\mu$m. 
\\

 For HD135344, the disk parameters deduced from fitting only the SED of HD135344 were close to those which allows to fit both the SED and our observations. 

  For CQTAU, in addition to previous studies fitting only the SED, we had to take into
  account that the disk is quite extended at 20 $\mu$m (about 300 AU)
  and is observed with an inclination of 33$^{\circ} \pm$ 5. The fast increase of the disk emission in 
  the [10-30] $\mu$m range can be modeled only with a flared disk (with H$_{p}^{out}/R_{out} > 0.1$). 
  In the framework of our modeling, we found that the only manner to obtain simultaneously
  a quite low far-IR excess (as seen in the spectrum) with the
  observed extended 20 $\mu$m emission is a low-mass disk of only 0.005 M$_{\odot}$ with a pressure scale height of 58 AU at 450 AU. The total flux in the infrared excess compared to the stellar flux is determinated by the fraction of the central star energy intercepted by the disk. This covering fraction is linked to H$_{p}^{out}/R_{out}$, where the disk geometry thickness is maximum. Here, this parameter is the same as the previous study (energy conservation) but for a different outer radius. That means that the disk is less flared than deduced earlier.
   We obtain for CQ~Tau a disk mass 10 times smaller than already found \citep{TESTI03, CHIANG01} and this mass only traces the small grains. The disk emission at mm wavelength is determined by the mid-plane grain and disk properties, and it is not affected by the nature of the surface dust. Consequently, in this paper, we have not tried to select the best parameters for the mid-plane dust to fit mm observations. Our underestimated flux at mm wavelength suggests that there must be big grains in the disk mid-plane of CQ~Tau to recover the measured flux at these wavelengths \citep{TESTI03}. 


HD~163296 is classified as a group II object
(interpreted as being surrounded by flat or a self-shadowed disk); it
should be in principle much more difficult to resolve the disk in the
mid-IR range.  Surprisingly, some extended mid-IR emission at 20
$\mu$m, although less prominent than for group I objects, is however
observed.
In the modeling, we used as first guess the parameter found in \citet{DD03},
who modelized the SED with a flared disk cut at 50 AU. Here, we modified 
the large-scale pressure scale-height in order to mimic a weakly flared
disk. The SED is well reproduced with a disk having little flaring (Fig~\ref{spectrum}
and Tab.~\ref{resultat_model}) while the extension at 20
$\mu$m (Fig.~\ref{model_profile}) constrains the disk to 
have a minimum outer radius of 200 AU with a disk mass of 0.01 $M_{\odot}$. 

\begin{table*}[ht]
\begin{minipage}{\hsize}
\setlength{\tabcolsep}{1.3mm}
\renewcommand{\arraystretch}{1.}
\renewcommand{\footnoterule}{}

\caption{Stellar properties and fit parameters. The stellar parameters
  of CQ~Tau are taken in \citet{CHIANG01}. For HD~135344 and HD~163296,
  the parameters are taken from \citet{MEEUS01} and \citet{VB05}. $H_{p}^{in}/R_{in}$ characterizes the puffed inner
  rim. $q$ is the power-law index of the scale height and $p$ that of the
  surface density. H$_{p}^{out}/R_{out}$ characterizes the flaring angle of the disk. The last column is the flux at 20.5 $\mu$m
  calculated from the modeled image.}\label{resultat_model}
\begin{tabular}{lcccccccccccc}
 \hline\hline
 Object   & distance & $M_{\star}$ & $T_{eff}$ & $R_{\star}$&   $R_{in}$ & H$_{p}^{in}/R_{in}$     & $R_{out}$ & H$_{p}^{out}/R_{out}$&  p  &  q &$i $& F(20.5 $\mu$m)\\
          & (pc) & ($M_{\odot}$)&  (K)        &  ($R_{\odot}$)&   (AU) &    &(AU) &   & &  &(degrees)& (Jy)  \\
 \hline
 \hline
HD~135344 &  84 &     1.3         & 6750            & 2.1  & 0.24 & 0.065  & 800\footnote{Disk size compatible with new measurement due to the re-evaluation of the distance.}  &0.21   &    0.8 & 9/7 & 60 & - \\
(DDN03) &      &  &                &       &      &        &       &       &        &        &   &\\
HD 135344 & 140 &1.3         & 6750            & 2.1  & 0.24 & 0.065  & 200  & 0.12    &  0.8 & 9/7 & 45 &5.7 \\
(in this study)&        &                &       &      &        &       &       &        &       &  & \\
\hline
HD 163296 & 122 &2.5          & 10500         & 1.7  & 0.45  & 0.033  &50   &  0.07   & 0.2 & 9/7    & 65 & -\\
(DDN03) &     &   &                &       &      &        &       &       &              &  & \\
HD 163296 & 100 &2.5          & 10500         & 1.7  & 0.45  & 0.033  &200   &  0.05     & 1. & 1.07    & 60 & 16.6\\
(in this study)&        &                &       &      &        &       &       &   &      &       &   &\\
\hline
CQ~Tau    & 100 & 1.7         & 7130          &  1.3 &  0.23 &  0.0002 &180 &  0.13   & 1.5   &  9/7  & - & -\\
(Chiang2001) &  &      &                &       &      &        &       &       &    &          &  & \\
CQ~Tau    &  100 &1.7         & 7130          &  1.9 &  0.23 &  0.018  &450 &  0.13       & 0.3   &  9/7  & 33 & 20\\
(in this study)&  &      &                &       &      &        &       &       &    &          & &  \\
\hline



\end{tabular}
\end{minipage}

\end{table*}

\begin{table*}[ht]
\begin{minipage}{\hsize}
\setlength{\tabcolsep}{1.3mm}
\renewcommand{\arraystretch}{1.}
\renewcommand{\footnoterule}{}
\caption{Effects of the free parameters of the model on the SED and spatial distribution of the 20.5 $\mu$m emission in the CAMIRAS image.}\label{influence_para}
\begin{tabular}{l|l|l}
 \hline\hline
 Parameter & Influence on SED & Spatial distribution at 20.5 $\mu$m  \\
 \hline
 $\Sigma_{0}$ & Influence on the total flux in the mid-IR   & No influence    \\
              & and more strongly for $\lambda \ge$ 60 $\mu$m &            \\
               &                                       &                \\
 $p$           & No influence in the mid-IR      & little influence       \\ 
               &                                       &                \\
 H$_{p}^{in}/R_{in}$ & modify contrast mid/near-IR & No influence     \\
                      & (structure inner rim/shadow) & (inner rim + shadow in the first pixel) \\
		                  &                                       &                \\  
$q$             &  modify the whole shape of the SED & large influence    \\
              &                                       &                \\
  H$_{p}^{out}/R_{out}$ & modifiy flux in the mid and far-IR & little influence \\
                &                                       &                \\
  $R_{out}$   & modify far-IR emission &  large influence    \\
  \hline
    \end{tabular}

\end{minipage}

\end{table*}
           
%


\section{Conclusions and future work}
We have shown the strength of mid-IR imaging to constrain the disk properties. 
There are still a very limited number of objects with extended emission spatially resolved and it is not yet possible to 
draw statistical conclusions about the spatial structure of disk around HAe stars. 
With the advance of mid-IR instruments on 8 meter class telescope, such as the VISIR \citep{LAGA04}
instrument available on the MELIPAL Very Large Telescope (VLT) at the European Southern 
Observatory (ESO), higher angular resolution will be available and 
the field will further develop; more quantitative studies will be possible and the goal of retrieving detailed disk surface density 
profiles from the observations should be achievable.\\
Observations should not be limited to the 20 $\mu$m atmospheric window; 
observations in the 10 $\mu$m atmospheric window are also very promising, 
especially for those HAe stars whose spectrum features the so-called PAH bands features at 3.3, 6.2, 7.7, 8.6, 11.2, 12.7 $\mu$m. 
PAH bands are attributed to vibrational relaxation of UV-pumped Polycyclic Aromatic
Hydrocarbon molecules containing about 50-100 carbon atoms
\citep{ALL89, PUGET89}. Their emission, as a function 
of the distance $r$ to the star, drops with a $r^{-2}$ power-law, much 
slower than thermal emission from large grains in thermal equilibrium. 
PAHs emission is thus a promising probe to study flaring disks 
at large distances from the star, with the good angular resolution achieved now on large ground-based
telescopes at 10 microns. Fitting by sophisticated models a combination of interferometric observations in the near- and mid-IR, which probe the inner-most disk regions, with single dish observations in the near and mid-IR which probe intermediate disk regions, is clearly the way to clear up the field in the next few years.  
\\

\clearpage{\pagestyle{empty}}

\newpage
 
\emph{Acknowledgments.} We are gratefully indebted to P. Masse, R. Jouan and M. Lortholary 
for their assistance with CAMIRAS instrument, A. Claret in efficiently supporting us in 
our observations, as well as to the staff of CFHT/Hawaii for their support during the observing 
runs.
CD wishes to thank J. Bouwman and E. Habart for very helpful discussions and advice.

\bibliographystyle{aa}

\bibliography{biblio_camiras}

 \end{document}